\documentclass{cs19proc}

\usepackage{kantlipsum}

\editors{G.~A. Feiden}
\publisher{Zenodo}
\conference{The 19th Cambridge Workshop on Cool Stars, Stellar Systems, and the Sun}
\conferencedate{2016}

\title{Relation between Brown Dwarfs and Exoplanets}
\author{Lauren Melissa Flor Torres$^{1}$, 
        Roger Coziol$^{1}$, 
        Klauss-Peter Schr\"oeder$^{1}$, 
        C\'esar A. Caretta$^{1}$ 
        and Dennis Jack$^{1}$}

\affiliation{$^{1}$ Departamento de Astronom\'ia, Universidad de Guanajuato, Guanajuato, M\'exico }

\shorttitle{Brown Dwarfs vs. Exoplanets}
\shortauthors{L. M. Flor Torres et al.}

\abs{One of the most debated subjects in Astronomy since the discovery of exoplanets is how can we distinguish the most massive of such objects from very-low mass stars like Brown Dwarfs (BDs)? We have been looking for evidences of a difference in physical characteristics that could be related to different formation processes. Using a new diagnostic diagram that compares the baryonic gravitational potential (BGP) with the distances from their host stars, we have classified a sample of 355 well-studied exoplanets according to their possible structures. We have then compared the exoplanets to a sample of 87 confirmed BDs, identifying a range in BGP that could be common to both objects. By analyzing the mass-radius relations (MRR) of the exoplanets and BDs in those different BGP ranges, we were able to distinguish different characteristic behaviors. By comparing with models in the literature, our results suggest that BDs and massive exoplanets might have similar structures dominated by liquid metallic hydrogen (LMH).}

\begin{document}

\maketitle

\section{Introduction}

The most accepted interpretation of Brown Dwarfs (BDs) is
that they are failed stars \citep{Cushing2014}, because, although 
it is assumed they formed like stars, their masses are too small to 
permit the fusion of hydrogen in their nucleus. This characteristic 
allows to separate BDs from main sequence stars based on their masses: 
because a star must reach a critical mass to be able to burn its 
hydrogen, which varies from $0.07\ M_\odot$ for  solar metallicity 
to $0.09\ M_\odot$ for lower metallicities \citep{Burrows2001}, any star 
with a mass  $< 70\  M_J$ (where  $M_J$ is the mass of Jupiter) is a  
BD \citep{Bate2006}. 

However, determining a lower mass limit for a BD is more difficult. 
In practice, the consensus to adopt the critical mass for the fusion of 
deuterium, which is around $13\ M_J$ \citep{Bate2006}, is arbitrary, because theoretically
the lowest mass a BD could have may be just a few $M_J$ 
\citep{Larson1969,Rees1976,Silk1977a,Silk1977,Boss1988}. 
Interestingly, this mass is also typical of massive exoplanets, and, since there is no obvious upper-mass limit for an exoplanet, 
hence, persists the problem of distinguishing between to two objects.

In this poster, using a large sample of ``well-studied'' exoplanets, 
and comparing with a large sample of ``confirmed'' BDs available in 
the literature, we probe a mass range common to both classes of 
objects, looking for evidence of a difference between their respective 
physical structures, as reflected by their mass-radius relations (hereafter MRRs). 

Our study concentrates on two questions: 1) At what mass boundary should we expect to see a variation in the 
MRR that would be consistent with a difference of structure between 
exoplanets and BDs?  2) Is there a special intermediate mass range where these two classes 
of objects are likely to overlap in mass? In particular, we propose a lower-mass limit for BDs based on the Self-Gravitating (SG) limit, which marks the moment the self-gravity of matter begins to affect significantly the structure of a body \citep{Padmanabhan1993}.

In addition to the MRR, the distance of a planet from its host star 
could also reveal something about its formation process 
\citep{Lissauer1993}. For the exoplanets, this last parameter is 
fundamental to identify Hot Jupiters \citep{Johnson2009}, while for the BDs, this parameter can be used to test the
 ``BD's desert'' hypothesis, which according to some authors  \citep[e.g.,][]{Grether2006}
might be related to different formation processes for exoplanets and BDs.

\section{Samples}

Our sample of exoplanets consists of 355 entries in the 
latest issue of the transiting planets catalog available at 
TEPCat\footnote{http://www.astro.keele.ac.uk/jkt/tepcat.html}, and can be considered as an upgraded version of 
the sample of well-studied exoplanets used previously in the study of 
\citet{Hatzes2015}. Note that because these exoplanets are detected by 
the transit method, their uncertainties on the inclination of their 
orbits, $i$, are relatively low \citep{Winn2010,Koch2010,Batalha2014}, 
which reduces significantly the uncertainties on their masses, 
$M=M_{planet} \sin i$. In our sample, the median uncertainties are 6\% for 
the masses and 5\% for the radius.

Our sample of BDs is composed of 87 objects selected from the upgraded 
compilation produced by \citet{Johnston2015}, which is based on published 
data. For all the BDs in our sample, we double-check their classification as 
BDs using SIMBAD. Although all these BDs have a mass and 
radius determined, only 37 have a distance estimate from a companion 
star. Of the remaining 50 BDs that do not have a distance reported in our list, 
14 are part of a binary system with another BD in our list (for which we have the distance), whereas 36 are genuine
isolated objects, which already makes them different from exoplanets.  

\section{The baryonic gravitational potential (BGP) diagram and the Self-Gravitating (SG) limit}

\begin{figure*}[ht]
\centering
\includegraphics[width=0.85\linewidth]{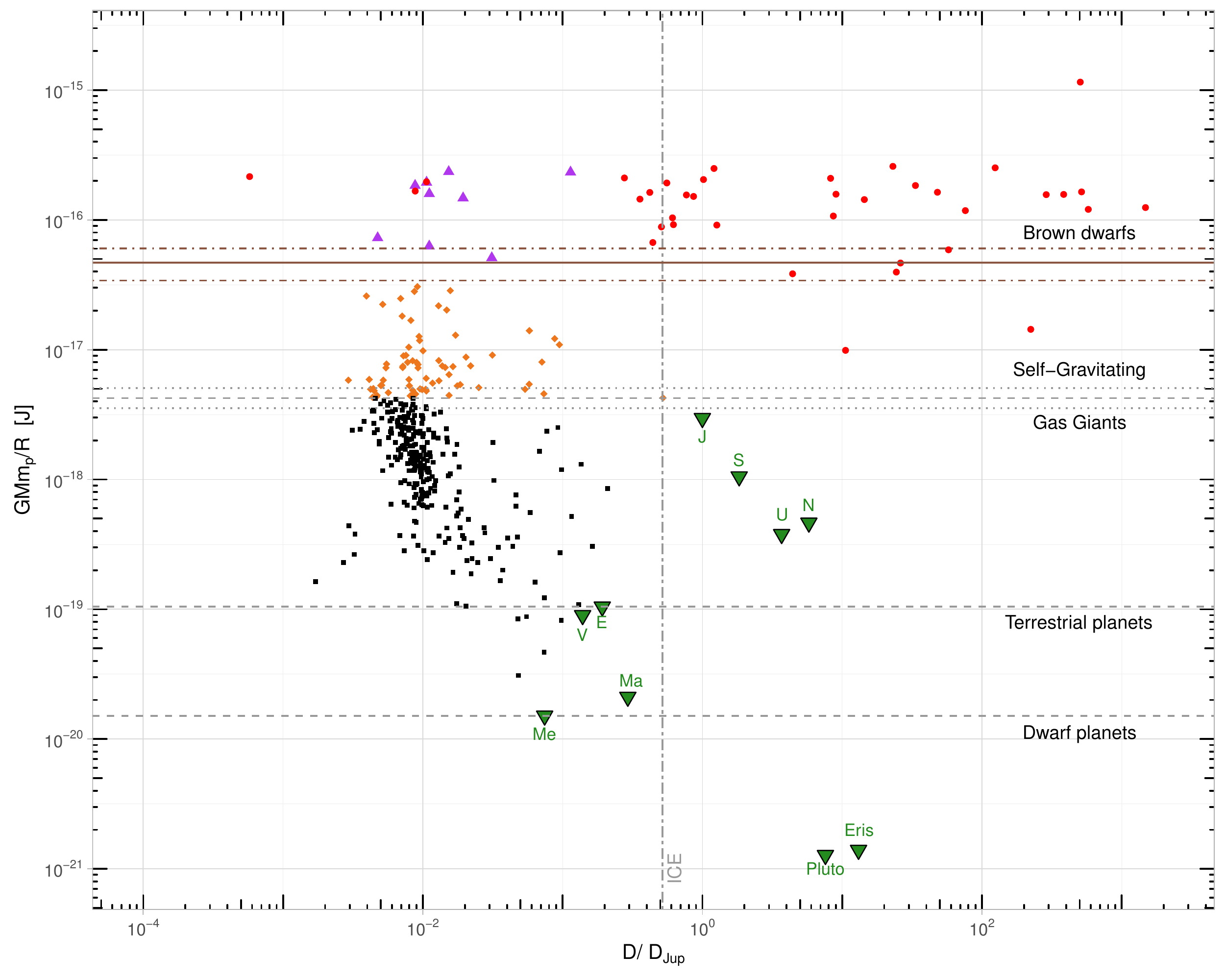}
\caption{The BGP diagram for exoplanets and BDs: nSGEs (black squares), SGEs (orange diamonds),  and BDs (red solid dots); the SGEs falling in BD region are identified by purple triangles. The inverted green triangles correspond to the positions occupied by the different kinds of planets in the solar system. The position of the ice line (or water "snowline") in the solar system (vertical dot-dash line) is also indicated.}
\label{fig:fig1}
\end{figure*}

To compare the exoplanets with the BDs, we combine the mass and radius 
into one physical parameter: the baryonic gravitational potential (BGP), which is defined as the gravitational 
potential energy of a body, divided by the number of 
its nucleons, $N$. Assuming the mass is $M = N m_p$, where $m_p$ is the 
mass of a proton, the BGP is thus equal to:
\begin{equation}
{\rm BGP} = \frac{V_G}{N} = \frac{G M m_p}{R} \propto \frac{M}{R}
\end{equation}
Note that since the BGP\ $ \propto M/R$, this parameter can be taken as 
a first order approximation for the MRR.  

In Figure~1 (hereafter, the BGP diagram) we compare for the exoplanets (black squares and orange diamonds) and BDs (red solid dots) the BGP and distances from their companion stars, as normalized by the distance of Jupiter from the sun (${\rm D}/{\rm D_{Jup}}$). The BGP diagram is separated in four zones, synonymous with different physical structures. The upper zone is defined by the lower mass limit of $13\ M_J$ for the burning of deuterium in BDs. 

Most of the exoplanets in our sample are Hot Jupiters, which is consistent with the well-known observational biases related to the detection methods. A few exoplanets are located above the deuterium-burning limit, while a few BDs are below this limit, suggesting that the deuterium-burning criterion does not allow a clear distinction between these two objects. Also, as observed by \citet{Santerne2016}, many BDs in our sample are found at a distance nearer than Jupiter from the Sun, contradicting the BD's desert hypothesis. 

Therefore, although the majority of the exoplanets and BDs occupy different regions in the BGP diagram, their separation in terms of physical structures is still somewhat ambiguous.   

\begin{figure*}[ht]
\centering
\includegraphics[width=0.85\linewidth]{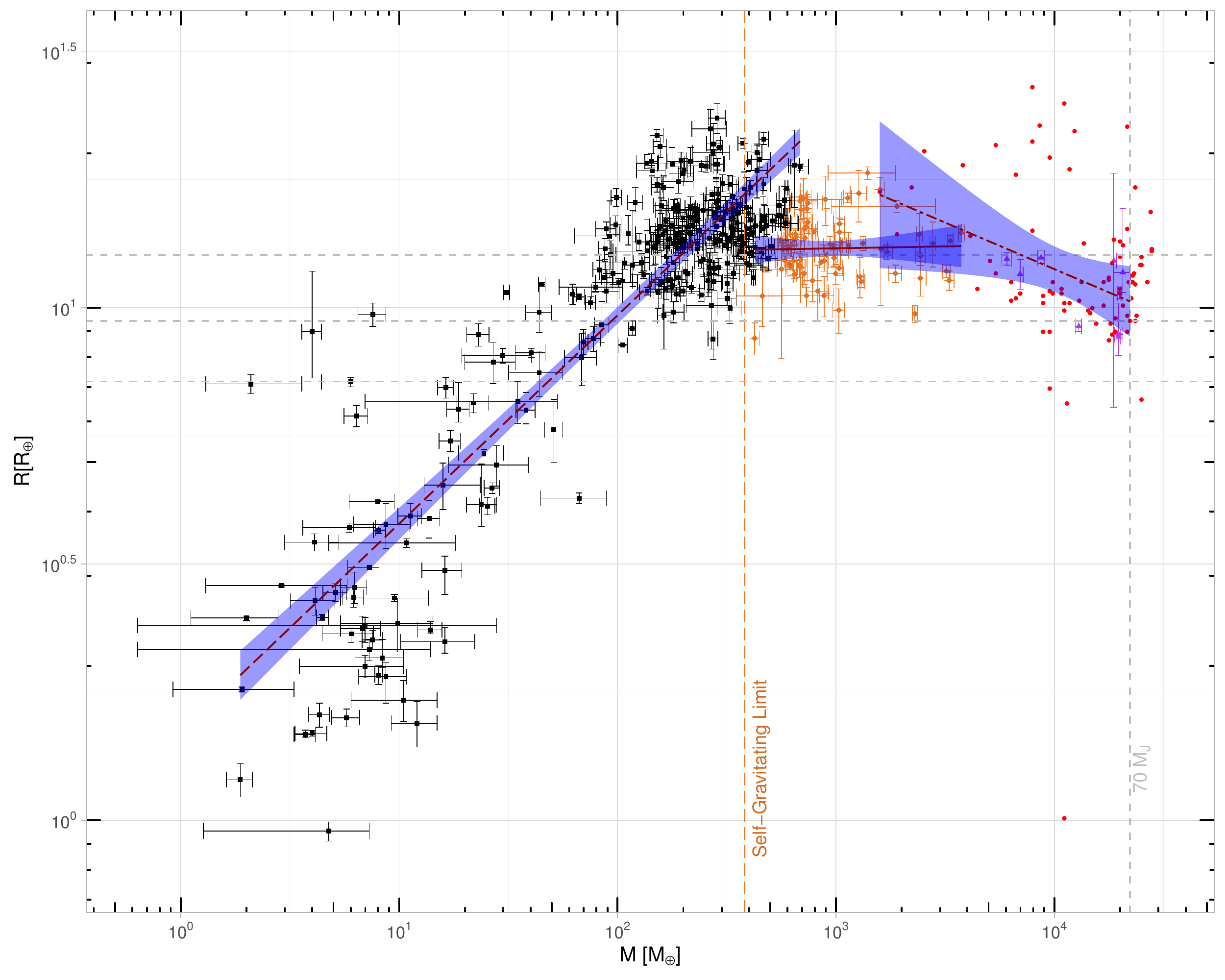}
\caption{Comparing the MRRs of exoplanets and BDs. The MRRs are traced with their corresponding 95\% confidence intervals. The symbols are the same as in Fig. 1. Also shown are the critical mass at the SG limit and the upper mass limit $70M_J$ for BDs.}\
\label{fig:fig2}
\end{figure*}

\begin{figure*}[ht]
\centering
\includegraphics[width=0.85\linewidth]{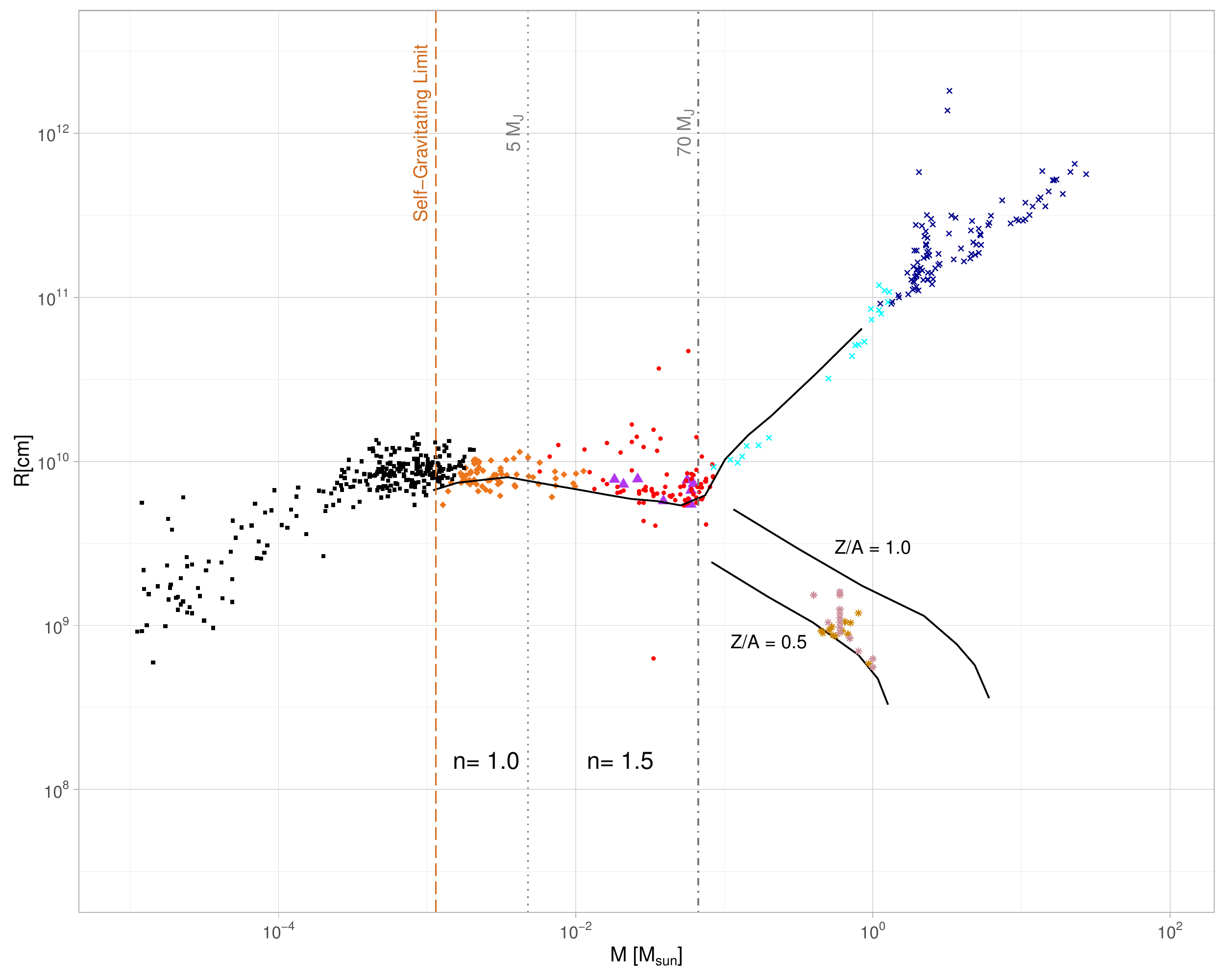}
\caption{Model of BDs formed of LMH. The unsignificant change of radius of the SGEs is due to a change of the polytropic index, n, below M = 5 $M_{J}$ from 1.5 to 1.0. Just at the point where the main sequence thermonuclear burning starts, around $70M_{J}$, there is a bifurcation: the MRR for the very low-mass stars (VLM, in light blue) becoming positive as they evolve towards the main sequence (dark blue x), while the MRR for WDs (white dwarfs, brown squares) is still negative. Two MRR for WDs, with different hydrogen richness (Z/A = 1.0 and Z/A = 1.5), are also represented. }
\label{fig:fig3}
\end{figure*}

Based on the SG limit, we separated the gas-giant exoplanets in Self-Gravitating (SGE; orange diamonds) and non Self-Gravitating (nSGE; black squares).   
The BGP for the SG limit is defined by a critical mass, $M_c$, and critical radius, $R_c$ \citep{Padmanabhan1993}. 
At the SG limit, the maximum number of baryons that an object can contain, $N_{max}$, is equal to:
\begin{equation}
N_{max}  = (\alpha/\alpha_g)^{3/2} \sim 1.38\times 10^{54}
\end{equation}
where $\alpha$ is the fine-structure constant and $\alpha_g$ the equivalent constant for gravity.
This corresponds to the critical mass:
\begin{equation}
M_c = N_{max} m_{p} \sim 2.31 \times 10^{27} {\rm kg} \sim 1.2 M_J
\end{equation}
 Then, assuming the radius of such object is $R =  N_{max}^{1/3} a_0$, where 
$a_0$ is the radius of Bohr, we obtain the critical radius: 
\begin{equation}
R_c = 6 \times10^7 {\rm m} \sim 0.84 R_J
\end{equation}

Note that although both $M_c=1.2 M_J$ and $R_c=0.84 R_J$ are typical values 
for massive exoplanets, the critical mass is also comparable with the 
theoretical lowest mass expected for a BD, while the critical radius
is consistent with their observed mean radius 
\citep{Burgasser2008,Basri2006,Sorahana2013}.  

\section{Mass-Radius Relation (MRR)}

According to \citet{Padmanabhan1993}, the MRRs of 
bodies with different structures would be expected to change abruptly at the 
SG limit, from a positive MRR below the SG limit, to a negative one 
above it, which may help distinguishing between exoplanets and BDs. 
Indeed, this is what we observe in Figure~2, where we compare the MRRs for the exoplanets and BDs: the nSGEs show a positive MRR while the BDs show a negative one (see Table~1). On the other hand, the SGEs show a relation where the radius does not increase with the mass. Note that this characteristics was also observed by \citet{Hatzes2015}, although these authors did not offered any physical explanation for this behavior. 

\begin{table}[t]
\centering
\caption{Linear regression in log, $(R/R_{\oplus}) = 10^{b}\times (M/M_{\oplus})^a$, and their coefficients of correlation $r^2$
}
\begin{tabular}{l c c c}
\noalign{\smallskip}\hline\hline\noalign{\smallskip}
\textbf{Sub-samples} & \textbf{a} & \textbf{b} & $r^2$\\
\noalign{\smallskip}\hline\noalign{\smallskip}
nSGE & $+0.41 \pm 0.01$ & $0.17 \pm 0.03$ & 0.785 \\
SGE & $+0.01 \pm 0.03$ & $1.09 \pm 0.10$ & 0.001 \\
BD & $-0.18 \pm 0.08$ & $1.60 \pm 0.33$ & 0.069 \\
\noalign{\smallskip}\hline
\end{tabular}
\end{table}

For the SGEs, we interpret the radius that shows NO significant change as the mass increases as evidence for the presence of a dominant liquid metallic hydrogen (LMH) envelop \citep{Wigner1935,Hubbard1997,Dalladay-Simpson2016}: this is due to the very low compressibility of LMH \citep{Hubbard1997}. 

That gas-giant planets, like Jupiter and Saturn in the solar system, have a LMH envelope was suspected by many 
authors since a very long time \citep[see][and reference therein]{Burrows1993}. But, one would not expect to observe evidence for such envelopes. This is because, although the LMH layer could constitute 50\% to 85\% of the 
mass of a gas-giant planet, this layer would generally be hidden below a rich envelope of hydrogen gas. What we think could have happened, therefore, is the following. As the mass of a gas-giant exoplanet increased above the critical mass, $M_c$, the self-gravity of 
matter became more important, the pressure increased and most of 
the hydrogen in the outer gas envelope changed phase, transforming into 
LMH. Alternatively, since most of these exoplanets are Hot Jupiters with high eccentricities, they might have lost their outer envelop of gas when passing near their stars, revealing their underlying LMH envelopes. 

However, based on the LMH interpretation there is still another alternative, which is that above the SG limit, objects are really BDs. In Figure~3 we compare our data with the predictions made by such a model, as developed by \citet{Burrows1993}. In this model BDs are formed at 99.9\% of 
LMH, this percentage decreasing as the mass of the star decreases, down to the SG limit. Below $5\ M_J$, the Coulomb correction competes 
with the degeneracy component, and the polytropic index, $n$, changes 
from 1.5 to 1.0, making the radius independent from the mass. 
Note that, according to this model, even above $5\ M_J$, 
when $n = 1.5$ and $R \propto M^{-1/3}$, the dependence of the radius on 
the mass would be weak, consistent with the low coefficients of correlation we observed.  

\section{Conclusions}

\begin{itemize}

\item We conclude that the unsignificant change of radius of exoplanets above the SG limit is the characteristic signature of objects formed by LMH.

\item As for the nature of these objects we propose that they could be either giant gas planets with a dominant layer of LMH, more massive than what is assumed to exist in Jupiter and Saturn, or genuine very low-mass BDs.

\end{itemize}

\section*{Acknowledgments}
{L.M.F.T. acknowledges the ``Direci\'on del Campus Guanajuato'' 
(Folio 10040) of the University of Guanajuato, for special 
support and travel grants. 
L.M.F.T. and K.-P.S. both acknowledge travel support by the bilateral 
Conacyt-DAAD PROALMEX programme (No. 207772). 
C.A.C. thanks finantial support from DAIP's ``Convocatoria
Institucional'' project \#627/15.
Our study made use of the Johnston's Archive, which is copyrighted by 
Wm. Robert Johnston. This research has made use of the SIMBAD database,
operated at CDS, Strasbourg, France.}

\bibliographystyle{cs19proc}
\bibliography{cs19_flortorres_etal.bib}

\end{document}